# A BAYESIAN MODEL COMMITTEE APPROACH TO FORECASTING GLOBAL SOLAR RADIATION

Philippe Lauret
Hadja Maïmouna Diagne
Mathieu David
PIMENT – University of La Reunion
97715 Saint Denis Cedex 9
hadja.diagne@univ-reunion.fr
mathieu.david@univ-reunion.fr
philippe.lauret@univ-reunion.fr

Auline Rodler
Marc Muselli
Cyril Voyant
CNRS UMR SPE 6134
University of Corsica
20250 Corte, France
muselli_m@univ-corse.fr

ABSTRACT

This paper proposes to use a rather new modelling approach in the realm of solar radiation forecasting. In this work, two forecasting models: Autoregressive Moving Average (ARMA) and Neural Network (NN) models are combined to form a model committee. The Bayesian inference is used to affect a probability to each model in the committee. Hence, each model's predictions are weighted by their respective probability. The models are fitted to one year of hourly Global Horizontal Irradiance (GHI) measurements. Another year (the test set) is used for making genuine one hour ahead (h+1) out-of-sample forecast comparisons. The proposed approach is benchmarked against the persistence model. The very first results show an improvement brought by this approach.

## 1. INTRODUCTION

Solar radiation forecasting is of great importance for many applications such as for instance improved PV grid integration. More precisely, in order to increase the integration of renewables into non-interconnected electricity grids (such as insular grids), accurate forecasts at various time steps or forecast horizons are needed.

The solar radiation sequence can be seen as a time series, and therefore one can build statistical models to capture the underlying random process and predict the next values. Several statistical techniques can be employed to forecast the solar radiation time series. The spectrum of methods ranges from linear models like the Autoregressive Moving Average (ARMA) model to non linear models like Neural Networks (NNs).

In the realm of time series forecasting, some authors (1), (2) proposed to combine the predictions of ARMA and NN models (hence giving a so-called hybrid model) in order to improve the forecasting accuracy achieved by either of the models used separately.

However, to our best knowledge, it must be stressed that the combination consists in adding simply the linear part to the nonlinear part. In others words, the same weight is given to each model in the committee.

In this work, we propose to combine the two models (ARMA and NN) and to weight their predictions by a measure of their respective performance. The Bayesian inference is used to affect a probability to each model in the committee. Hence, each model's predictions are weighted by their respective probability.

The models are fitted to one year of hourly Global Horizontal Irradiance (GHI) measured at the site of Ajaccio (Corsica Island, France). In a second step, the one step ahead (h+1) predictions of the models are calculated with another year of hourly GHI. The first results show an improvement brought by this approach.

## 2. CONTEXT OF STUDY

In this survey, we focus on the forecasting of global horizontal irradiance (GHI) at a hourly time step. The data used to build the models are GHI measured at the meteorological station of Ajaccio (Corsica Island, France, 41°550N, 8°440E). Measurements are available on an hourly basis from January 1998 to December 2008.

A key feature in the identification of linear models like ARMA

models is the data transformation that is often needed to make the time series stationary. Stationarity means that the statistical characteristics of the time series such as the mean and the autocorrelation structure are constant over time (3). In this survey, as the solar radiation series is not stationary, we used the simplified Solis clear sky model (4) in an attempt to obtain a stationary hourly solar series. More precisely, we obtained a deseasoned series by applying the following data transformation:

$k_{cls} = \dfrac{I_g}{I_{g,clearsky}}$ (where $I_g$ is the global irradiance, $I_{g,clearsky}$ is the output of the Solis clear sky model and $k_{cls}$ is the so-called clear sky index). This transformation makes use of the fact that the global irradiance $I_g$ can be decomposed into a deterministic clear sky component and a stochastic cloud cover component.

One may notice however that this transformation is not optimal (i.e. the time series may still exhibit some heteroscedasticity) and one has to apply some more data transformations like differencing to remove the trend and/or stabilize the variance. Another possibility is to make use of integrated ARMA (ARIMA) models (3) in order to treat non-stationary series. This type of model will be investigated in future work.

3. ARMA model

In an ARMA model (3), the future value of a variable namely $y_t$ is assumed to be a linear combination of several past observations and random errors i.e. :

(1) $\quad y_t = \phi_0 + \sum_{i=1}^{p} \phi_i y_{t-i} + \varepsilon_t - \sum_{i=1}^{q} \theta_i \varepsilon_{t-i}$

where $\{y_t\}$ is the time series of interest and $\{\varepsilon_t\}$ is a white noise series. The model's parameters are the $\theta_i (i=1,2,\cdots,q)$ and $\phi_i (i=0,1,2,\cdots,p)$. The integers $p$ and $q$ are called orders of the model and one key challenge in the building of ARMA models is to determine the appropriate model orders. Several criteria are proposed to select the best orders. Among the latter, the Bayesian Information Criterion (BIC) is usually preferred. In the following, we will use the notation ARMA(p,q) to denote the model given by Eq. (1) and the series $\{y_t\}$ will represent the clear sky index time series i.e. $\{k_{cls}\}$

4. A NN APPROACH TO SOLAR RADIATION FORECASTING

The use of Neural Networks (NNs) is particularly predominant in the realm of time series forecasting. Indeed, the availability of historical data on the meteorological utility databases and the fact that NNs are data driven approaches capable of performing a non-linear mapping between sets of input and output variables make this modeling tool very attractive. NNs are able to approximate any continuous function at an arbitrary accuracy, provided the number of hidden neurons is sufficient. However, it is necessary to match the complexity of the NN to the problem being solved. The complexity determines the generalization capability (measured by the test error) of the model since a NN that is too complex will give poor predictions. In the NN community, this problem is called overfitting. Several techniques like pruning or Bayesian regularization (5) can be employed to control the NN complexity.

Several NN architectures are proposed in the literature but for time series forecasting, the most common architecture is the so-called feedforward architecture with one single layer of hidden neurons. For a NN having $h$ hidden neurons, the relationship between the output $y_t$ and the inputs $y_{t-1}, y_{t-2}, \cdots, y_{t-p}$ has the following form:

(2) $\quad y_t = \alpha_0 + \sum_{j=1}^{h} \alpha_j f\left( \beta_{0j} + \sum_{i=1}^{p} \beta_{ij} y_{t-i} \right)$

where $\alpha_j$ and $\beta_{ij}$ are the model's parameters (also called weights in the NN terminology). The function $f(x)$ associated with an hidden neuron (or hidden unit) is usually the tangent hyperbolic function i.e. $f(x) = \dfrac{e^x - e^{-x}}{e^x + e^{-x}}$.

Thus, the neural network is equivalent to a nonlinear autoregressive (AR) model for time series forecasting problems.

At this point, it must be noted that the number of parameters for a single-layer is given by $(p+2)*h+1$ and this number is typically much larger than in linear time series models. One may notice that this fact may render the use of in-sample criterion like BIC innapropriate for the selection of the best NN model. However, in this work, we used the Bayesian regularization method depicted in (6) in order to overcome the overfitting problem. In the following, we will use the notation

NN(p,h) to denote the model given by Eq. (2).

## 5. THE BAYESIAN COMMITTEE THROUGH BAYESIAN MODEL AVERAGING

Model averaging is a group of methods for combining predictions from several models. In addition to parameter uncertainty, the methods consider model uncertainty. More precisely, in model averaging, predictions of each model are weighted with factors related to model performance. Bayesian model averaging (BMA) is model averaging in a Bayesian framework (7). In BMA, the model's weights are given by the Posterior Model Probabilities (PMP). Before describing the computation of the PMPs, let us start with a brief description of Bayesian inference.

## 6. BAYESIAN INFERENCE

In the Bayesian context, a probability represents a degree-of-belief (or encodes a state of knowledge), that is, how likely something is to be true based on all the relevant information at hand. As Bayesian probability theory does not define a probability as a frequency of occurrence but rather as a reasonable degree of belief, it is possible to assign probabilities to propositions such as "The probability that parameter $\theta_k$ had value x when data was taken". In other words, in the Bayesian framework, questions of the form : "What is the best estimate of a parameter one can make from the data and prior information?" make perfect sense.

### 6.1 Bayesian parameter estimation

The first level of Bayesian inference concerns parameter estimation. More precisely, assume that (among a set of $K$ possible models), model $M_k$ has a vector of $m_k$ parameters $\Theta_k = \left(\theta_{1_k}, \theta_{2_k}, \cdots, \theta_{m_k}\right)$. Bayesian inference deals with the estimation of the values of $m_k$ model parameters about which there may be some prior beliefs. These prior beliefs can be expressed as a probability density function (pdf) called prior, $p(\theta_k | M_k)$ and may be interpreted as the probability placed on all possible parameter values before collecting any new data. The dependence of the $n$ observations (or measurements) $D = (d_1, d_2, \cdots, d_n)$ on the $m_k$ parameters can also be expressed as a pdf: $p(D | \Theta_k, M_k)$, called the likelihood function. The latter is used to update the prior beliefs about $\Theta_k$, to account for the new data $D$. This updating is done through Bayes's theorem:

$$(3) \quad p(\Theta_k | D, M_k) = \frac{p(D | \Theta_k, M_k) p(\Theta_k | M_k)}{p(D | M_k)}$$

where $p(\Theta | D, M_k)$ represents the posterior pdf and expresses the values of the parameters after observing the new data. In other words, the prior is modified by the likelihood function to yield the posterior. As mentioned above, a major difference between Bayesian and frequentist (or classical) methods is that the Bayesian inference offers a framework (through the use of prior information) to continuously update our posterior beliefs.

The term $p(D|M_k) = \int p(D|\theta_k, M_k) p(\theta_k | M_k) d\theta_k$ is called the evidence term or marginal likelihood. We will see below that the marginal likelihood plays an important role in BMA.

### 6.2 Bayesian model selection

At the second level of inference, the problem consists in inferring which model is most plausible given the data. The posterior probability of each model is as follows:

$$(4) \quad p(M_k | D) = \frac{p(D | M_k) p(M_k)}{p(D)}$$

As mentioned above, $p(D | M_k)$ is the marginal likelihood for model $M_k$. The quantity $p(M_k)$ represents a prior belief for model $M_k$. If we have no particular reason to prefer one model over another, then we would assign equal priors to all models. Since the denominator does not depend on the model, one can see that the different models are ranked according to the evidence term $p(D | M_k)$.

### 6.3 Bayesian Model Averaging

Instead of picking the most probable model, one can sum over all the models. As an illustration, consider a set of $K$ models. The model ensemble posterior distribution of a quantity $Q$ (for instance the future model predictions using new input data) given the data D, is:

$$p(Q|D) = \sum_{k=1}^{K} p(Q|M_k, D) p(M_k|D) \quad (5)$$

where $p(Q|M_k, D)$ is the posterior distribution of $Q$ under model $M_k$ and data D and $p(M_k|D)$ is the posterior model probability or model weight.

Eq. (5) is a linear combination of the predictions made by each model separately where the weighting coefficients are given by the PMPs.

In the case of a set of $K$ discrete models and by assigning equal priors to all models, the PMP of model $M_k$ is given by:

$$p(M_k|D) = \frac{p(D|M_k)}{\sum_{i=1}^{K} p(D|M_i)} \quad (6)$$

As seen, PMPs are positive scalar and sum to unity. Model ensemble predictions are made by weighting the mean posterior predictions for each model by the model PMPs.

## 7. APPROXIMATING THE PMPS

In most cases, the computation of the marginal likelihood is analytically intractable. One possibility is to make use of the Laplace approximation that aims to find a Gaussian approximation to a probability distribution. This approximation produces accurate results provided the sample size is sufficient. From this previous approximation, we can derive the Bayesian Information Criterion (BIC) and we have:

$$BIC_{M_k} \simeq -2\ln(p(D|M_k)) \simeq \ln(\hat{\sigma}^2) + m_k \frac{\ln(n)}{n} \quad (7)$$

where $\hat{\sigma}^2$ is the estimate of the variance of the residuals and n is the number of observations. By using this approximation, the PMP of model $M_k$ can be re-written as (see (8) for more details):

$$PMP_{M_k} = p(M_k|D) \simeq \frac{\exp\left(\frac{-BIC_k}{2}\right)}{\sum_{i=1}^{K} \exp\left(\frac{-BIC_i}{2}\right)} \quad (8)$$

### 7.1 Bayesian committee

In our study, we restrict ourselves to a set of 2 models and we build a committee by weighting the predictions of the two models (ARMA and NN models) according to their PMPs, hence giving: $y_{t,committee} = PMP_{arma} \times y_{t,arma} + PMP_{nn} \times y_{t,nn}$

## 8. RESULTS AND DISCUSSION

Although eleven years of data were available, we deliberately used only two years to build and to test the models. Indeed, for the estimation of the models' parameters and for the computation of the PMPs, we used GHI from year 1998 (8760 hourly values). This dataset is called training set in the NN terminology (or in-sample dataset in the time series forecasting community). To assess the accuracy of the proposed models, we used GHI from year 2006. Again, in the NN community, this dataset is called test set (or out-sample dataset in the time series terminology).

The error metrics used for comparisons purpose are the classical Root Mean Square Error (RMSE in $Wh.m^{-2}$), Normalized Root Mean Square Error (nRMSE in %) and the Mean Bias Error (MBE in $Wh.m^{-2}$). Regarding the nRMSE metric, normalization is done with respect to mean ground measured irradiance of the year 2006. Table 1 gives the out-of-sample one-step-ahead (h+1) forecasting accuracy of the different models.

### 8.1 ARMA modelling

We investigated 110 different structures of ARMA models by varying the orders p from 1 to 10 and q from 0 to 10. An ARMA(2,1) exhibited the lowest BIC (see Table 1) which corresponds to a PMP of 0.4663 (see Table 2).

### 8.2 NN modelling

In a similar procedure, we tested several NN configurations and found that a NN(12,3) i.e with 12 hidden units and 3 lagged inputs exhibited a PMP of 0.5337 (see also Table 1). Notice again that the optimal NN complexity was found by using a Bayesian regularization method (see (6) for more details).

### 8.3 Model committee

The performance of the model committee (i.e. resulting from the combination of the NN model and the ARMA model) is given in Table 2. The model is governed by the following equation: $y_{t,committee} = 0.5337 y_{t,NN} + 0.4663 y_{t,ARMA}$

Fig. 1 and Table 2 show the improvement brought by the hybrid approach. Indeed, overall, the Bayesian model committee approach reduces the nRMSE of the persistence model by 9.5%. In addition, the hybrid approach led to a low biased model.

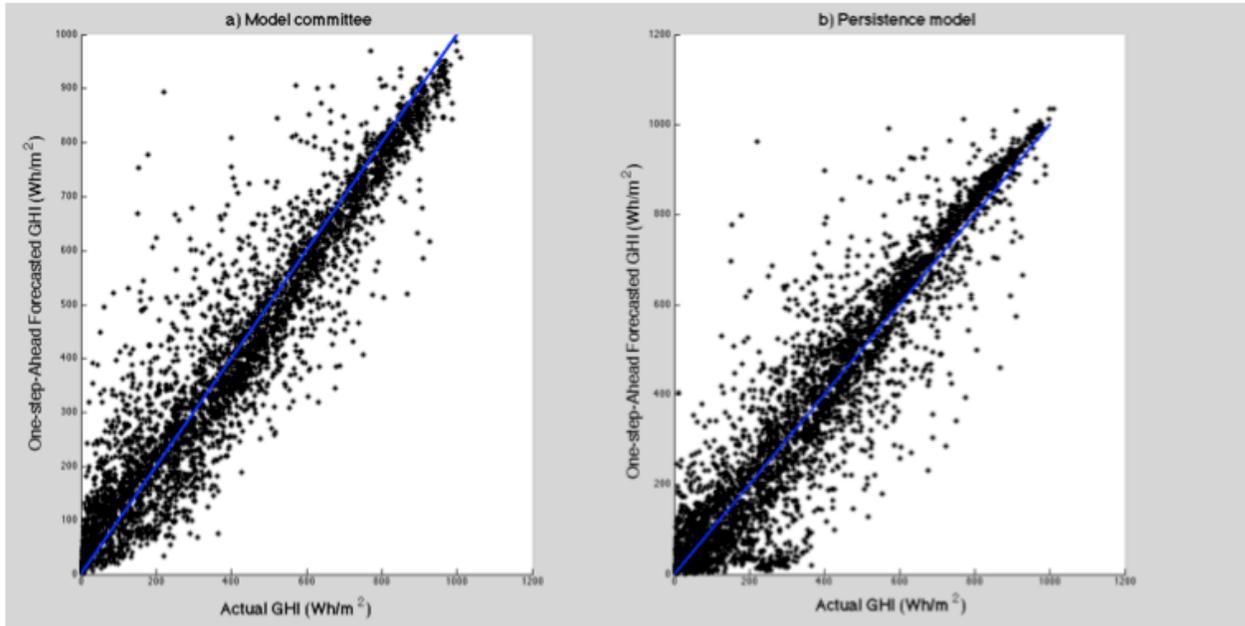

Fig. 1: Measured GHI vs. Forecast GHI (one-step-ahead prediction) a) Model committee b) Persistence

TABLE 1: OUT-OF-SAMPLE ACCURACY (OR TEST ERROR) OF THE DIFFERENT MODELS

| Molel | BIC | Exp(-BIC/2) | RMSE | nRMSE (%) | MBE |
|---|---|---|---|---|---|
| ARMA(2,1) | -2.33 | 3.20 | 95.92 | 25.01 | -4.08 |
| NN(12,3) | -2.59 | 3.66 | 88.65 | 23.10 | 3.06 |
| Persistence | -2.03 | 2.76 | 95.70 | 24.96 | 5.97 |

TABLE 2: PERFORMANCE OF THE COMMITEE

| PMP (NN) | PMP (ARMA) | RMSE | nRMSE (%) | MBE |
|---|---|---|---|---|
| 0.5337 | 0.4663 | 86.72 | 22.60 | -0.29 |

Fig.2 displays the forecasted irradiance in comparison to measured GHI for 4 days of year 2006. As seen and as expected, the forecast quality depends on the sky conditions. For clear sky conditions, the performances of the models are good but for days with clouds, performances naturally degrade. In other words, the hourly variation of the irradiance for days with variable clouds is not correctly predicted by the models. Future work will be devoted to a more in-depth analysis of the accuracy of the models for different sky conditions. It is planned also to assess the performance of the proposed method for different forecast horizons (say from h+1 to h+12). In addition, it must be stressed that the inputs of the models are only past observations of GHI. An improvement of the performances is expected by adding exogenous inputs to the model committee.

9. CONCLUSION

In this paper, we proposed a new methodology that consists in combining predictions of two solar forecasting models. The models' predictions are weighted according their corresponding Posterior Model Probabilities (PMPs). The first results show that the technique is a valuable approach for solar radiation forecasting as the hybrid methodology brings a clear improvement. The proposed statistical approach could complement (as a post-processing technique) and therefore increase the accuracy of existing forecasting methods based for instance on NWP methods. However, it must be pointed out that the goal of this work was to demonstrate the feasability of the proposed approach and that more effort must be devoted to the improvement of the forecasting accuracy. We can cite, among others, the addition of exogenous inputs (given for instance by total sky imagery) to the ARMA model or the NN model (hence giving the so-called ARMAX model or NNARX model).

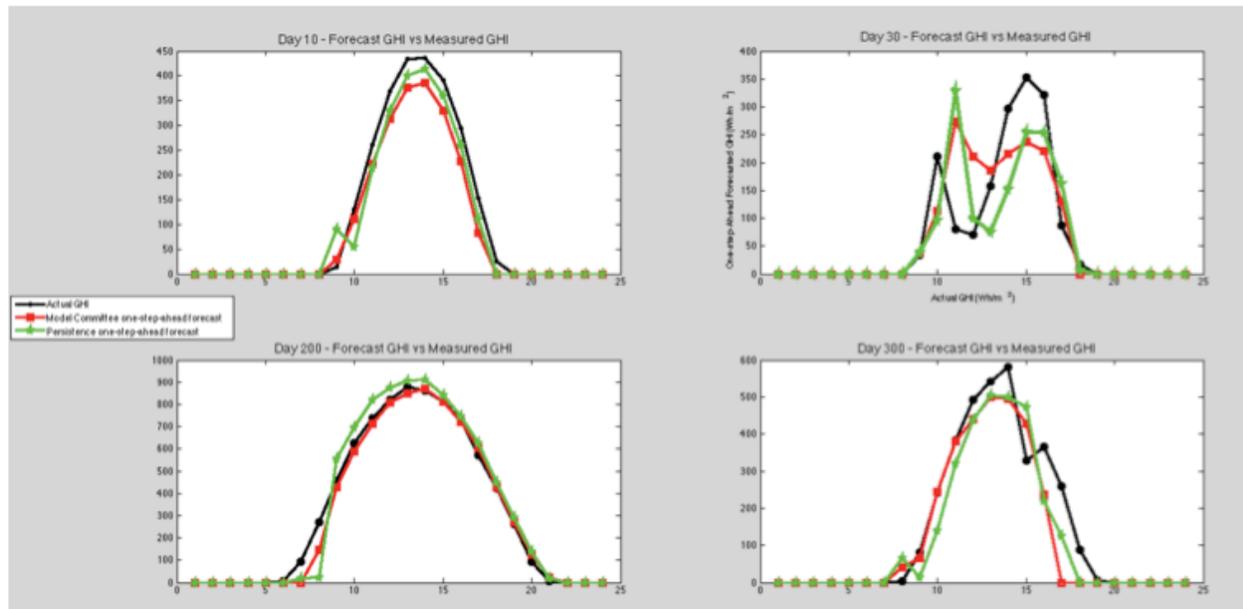

Fig. 2: Forecast of GHI : Model Committe (red) and Persistence (green) compared to measured GHI (black) for 4 days of year 2006